\documentclass[12pt]{article}
\input{epsfig.sty}
\newcommand{\beq}{\begin{eqnarray}}
\newcommand{\eeq}{\end{eqnarray}}
\begin{document}
\title{Sigma and Pion Decays of Mixed Heavy Hybrids}
\author{Leonard S. Kisslinger and Dara J. Krute\\
 Department of Physics, Carnegie Mellon University, Pittsburgh, PA 15213}
\date{}

\maketitle

\begin{abstract} We estimate the relative branching ratio of $\sigma$ to
2$\pi$ for the decay of the $\Psi'(2S)$, which we have found to be
a mixed charm and hybrid charmonium meson. The external field method
and correlators from a previous calculation are used to estimate these
decays. We find the $\sigma$ to 2$\pi$ branching ratio for  $\Psi'(2S)$ to 
the $J/\Psi(1S)$ decay to be 0.98, in agreement with an experiment at 
IHEP-BES.

\end{abstract} 
PACS Indices:14.40.Lb,13.25.Ft,12.38.Aw,11.55.Hx

\section{Introduction}

   For many years there have been experimental and theoretical studies of
the decay of the $\Psi'(2S)$ charmonium state. First, it was found that the
hadronic decays of the $\Psi'(2S)$ compared to the $J/\Psi(1S)$ were an order
of magnitude smaller than predicted by QCD perturbation theory\cite{markII}.
This was quite surprising, as for heavy quark systems the lowest order terms
should dominate. It is called the $\rho-\pi$ problem.
More recently, studies of final state interaction in the
2$\pi$ decay of the $\Psi'(2S)$ vs. the $J/\Psi(1S)$ were carried 
out\cite{gsc06}. This is directly related to our present work on $\sigma$ 
vs 2$\pi$ decay, since the $\sigma$ is a 2$\pi$ resonance (see, e.g., 
Ref \cite{zhou05}). Our approach, however, is quite different, since it
is based on the well-known strong coupling of the $\sigma$ to a gluon,
as discussed in detail below. See Ref \cite{gsc06} for references to earlier
theoretical studies related to 2$\pi$ decay of heavy quarkonium systems.

   In previous work\cite{lsk09} it was shown that the $\Psi'(2S)$ charmonium
and $\Upsilon(3S)$ bottomonium states are admixed hybrid and normal meson 
states, with approximately a 50\% mixture. For the $\Upsilon(nS)$ states 
this provides an explanation for the unusual two-pion decays\cite{hv06};
and for the $\Psi'(2S)$ this solves the famous $\rho-\pi$ problem.  Prior
to this, we had studied the $\Psi'(2S)$, also using the method of QCD sum 
rules, and found\cite{kpr08} that it was not a pure hybrid.
In the present work we estimate the decay of the $\Psi'(2S)$ to the 
$J/\Psi(1S)$ with a sigma or two free pions as a test of our mixed heavy 
quark hybrid meson theory.

   The operator which produces a mixed vector ($J^{PC}=1^{--}$) charmonium
state is 
\beq
\label{1}
        J^\mu &=& b J_H^\mu + \sqrt{1-b^2} J_{HH}^\mu \; ,
\eeq
with $b$ a constant and
\beq
\label{2}
          J_H^\mu &=& \bar{q}_c^a \gamma^\mu q_c^a  \nonumber \\
          J^\mu_{HH} &=&  \bar{q}_c^a\gamma_\nu G^{\mu\nu} q_c^a \; ,
\eeq
where  $J_H^\mu$ is the standard current for a $1^{--}$ charmonium state and 
$J_{HH}^\mu$ is the heavy charmonium hybrid current, with $q_c^a$ and 
$\gamma_\nu$ the charm quark field and the Dirac matrix, respectively.
$G^{\mu\nu}$ is
\beq
\label{3}
         G^{\mu\nu}&=& \sum_{a=1}^{8} \frac{\lambda_a}{2} G_a^{\mu\nu}
\; ,
\eeq
with $\lambda_a$ the SU(3) generator ($Tr[\lambda_a \lambda_b]
= 2 \delta_{ab}$) and $G_a^{\mu\nu}$ is the gluon field. 

  The method of QCD sum rules\cite{SVZ} uses a correlator, which we now 
define. Given the operator $J_A$, often called a current, which creates 
the state $|A>$ from the vacuum, $|>$, i.e., $J_A|> = |A>$, the correlator 
$\Pi_A(x)$ in coordinate space is defined by
\beq
          \Pi_A(x) &=& <|T[J_A(x)J_A(0)]|> \nonumber \; ,
\eeq
where T is the time-ordering operator.
For the investigation of a $1^{--}$ charmonium state, the correlators 
in coordinate space for a normal meson, $\Pi_H^{\mu \nu}(x)$, or a hybrid meson,
$\Pi_{HH}^{\mu \nu}(x)$, are
\beq
\label{4}
      \Pi_H^{\mu \nu}(x) &=& <|T[J_{H}^\mu(x) J_{H}^\nu (0)|> \nonumber \\
    \Pi_{HH}^{\mu \nu}(x) &=& <|T[J_{HH}^\mu(x) J_{HH}^\nu (0)|>  \; ,
\eeq
For the mixed charmonium state one uses the current $J^\mu$, Eq(1). 
The QCD sum rule method equates a dispersion relation for the correlator to
an operator product expansion, with a Borel transform to provide convergence.
In the present research we use correlators to obtain matrix elements
for decays of a mixed hybrid charmonium state that was found using QCD
sum rules, but we do not need the sum rules themselves.

In Ref\cite{lsk09} a QCD sum rule calculation was carried out using $J^\mu$
as the current for the correlator. It was shown that for both the $\Psi'(2S)$ 
and $\Upsilon(3S)$ states $b \simeq 0.7$, so that both of these states 
are approximately a 50\% admixture of normal and hybrid mesons.
In the present work we use these currents to estimate the branching ratios
for the decay of the $\Psi'(2S)$ to the $J/\Psi(1S)$ with either a sigma or
two pions, with the sigma being a broad 2-$\pi$ resonance with a mass of
about 400-600 Mev.

\section{$\sigma$ and 2$\pi$ Decays of the $\Psi'(2S)$}

   Using the external field method, the mechanism for the $2 \pi$ decay of
a normal component of the $\Psi'(2S)$ charmonium state to the $J/\Psi(1S)$ 
normal charmonium state is shown in Fig. 1. The correlator is defined with 
two currents, as in Eq(4), but the propagators for the $c$ or $\bar{c}$ are
in an external pion field, discussed in detail below. With this 
well-established theory there are no contributions with both pions coupled
to the $c$ or $\bar{c}$. In the figures the 1 at a vertex means a color
singlet, while an 8 is a color octet. Since a gluon is a color octet, a
hybrid meson consists of quarks in an octet coupled with a gluon to form a 
color singlet, as in the second line of Eq(2). $G-\pi$ coupling vanishes.

\begin{figure}[ht]
\begin{center}
\epsfig{file=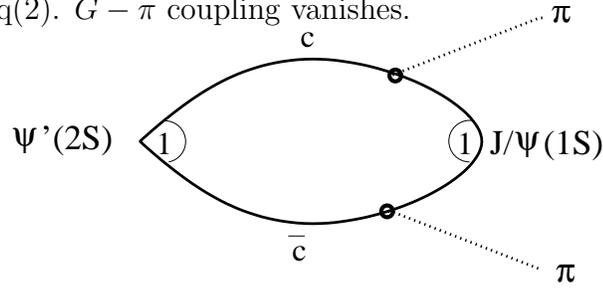,height=3cm,width=8cm}
\caption{Two free pion decay of a normal charmonium meson}
\label{Fig.1}
\end{center}
\end{figure}
 
 The mechanism for the $2 \pi$ decay of a hybrid charmonium to a 
charmonium meson is shown in Fig. 2, with the correlator defined using 
$J^\mu_{HH}$ and $J^\mu_H$ currents with $c$ and $\bar{c}$ external pion
field propagators.
The mechanism for $\sigma$ decay of a hybrid charmonium to a charmonium 
meson is shown in Fig. 3. The correlator differs from that of Fig. 2
in that there are no pion production operators, and the sigma is produced
by coupling to the gluon, G. There is little coupling of $\sigma$  to
quarks or antiquarks, which solves the $\Upsilon(nS)$ decay 
problem\cite{hv06,lsk09}.

\begin{figure}[ht]
\begin{center}
\epsfig{file=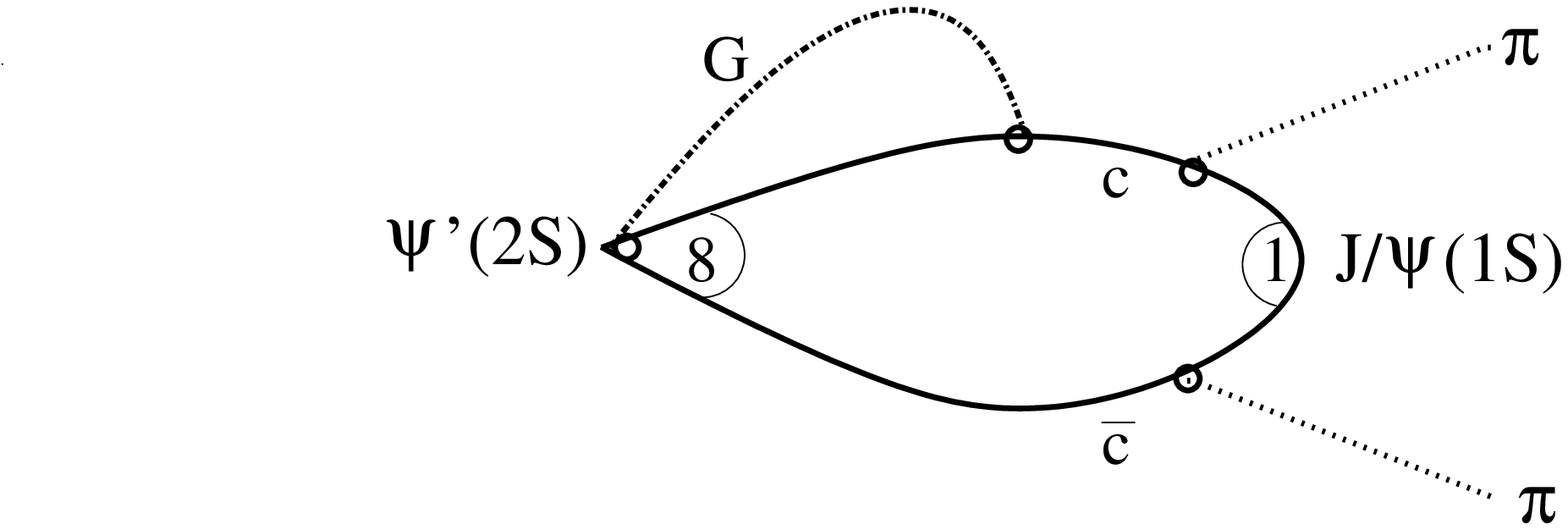,height=3cm,width=8cm}
\caption{Two-pion decay of a hybrid charmonium meson}
\label{Fig.2}
\end{center}
\end{figure}

\vspace{-1cm}

\begin{figure}[ht]
\begin{center}
\epsfig{file=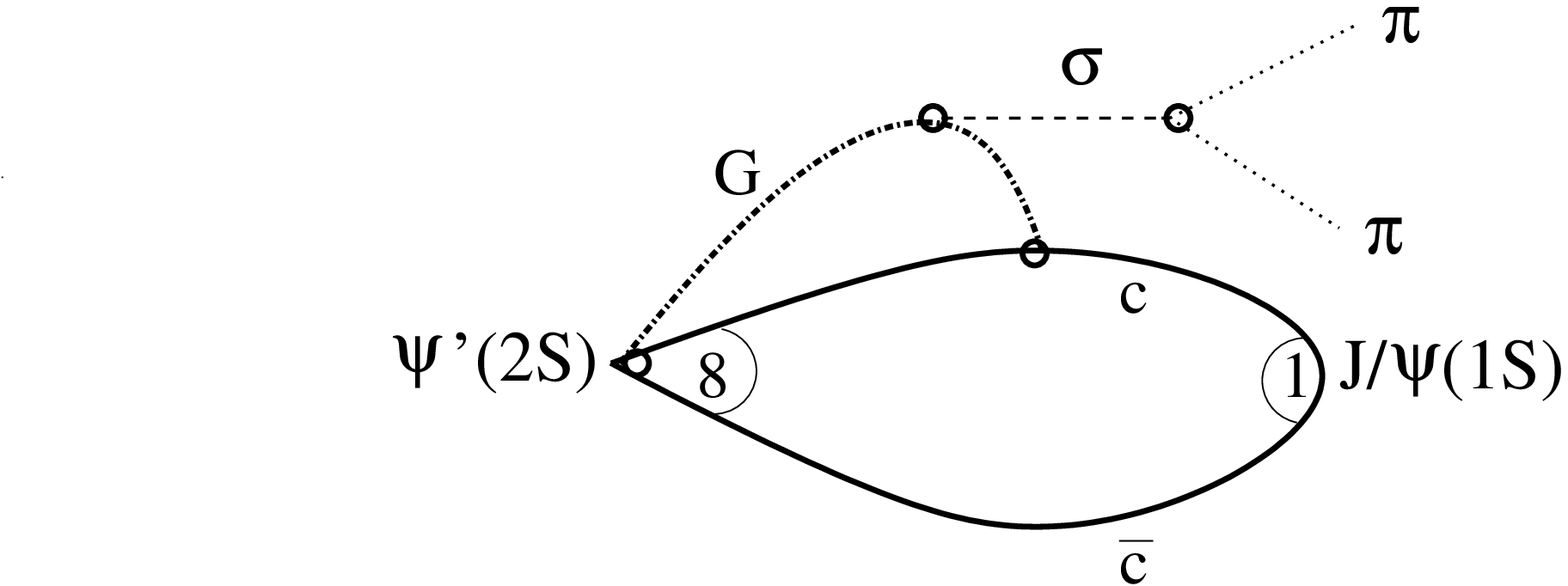,height=3cm,width=8cm}
\caption{Sigma decay of a hybrid charmonium meson}
\label{Fig.3}
\end{center}
\end{figure}
  
For the calculation of $\sigma$ and 2$\pi$ decays of a mixed hybrid state,
like the $\Psi'(2S)$, we use the external field method\cite{cpw85,hhk96}. 
The starting points 
are the diagrams used for the study of a mixed normal-hybrid 
state\cite{lsk09}
and the underlying diagrams for the decays shown in Figs. 1, 2, and 3. The 
coupling of mixed hybrid charmonium to normal charmonium is shown in Figs. 
4(a) and 4(b).

\begin{figure}[ht]
\begin{center}
\epsfig{file=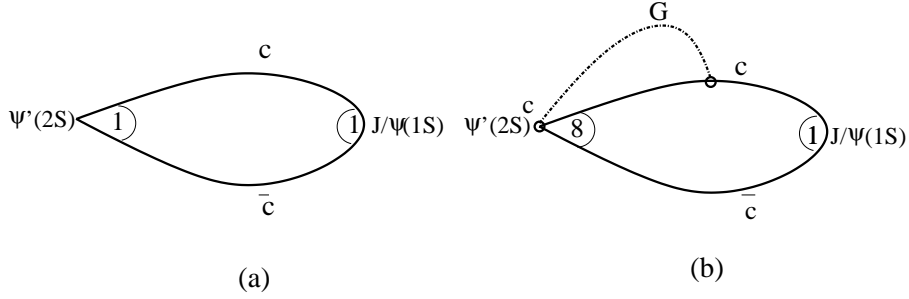,height=4cm,width=12cm}
\caption{Basic processes for mixed charmonium to couple to normal charmonium }
\label{Fig.4}
\end{center}
\end{figure}
Fig. 4(a) shows the mechanism for normal charmonium coupling, 
while Fig. 4(b) provides the basic mechanism for hybrid coupling to
normal charmonium. Using the usual notation of QCD sum rules these 
diagrams are the correlators $\Pi_{H}$ (Fig. 4(a)) and $\Pi_{H-HH}$
(Fig. 4(b)), respectively.

  The correlator corresponding to Fig. 4(a) is
\beq
\label{5}
   \Pi_{H}^{\mu \nu}(p) &=& 3 g^2 \int \frac{d^4 k}{(4 \pi)^4} Tr[S(k)
\gamma^\mu \gamma_5 S(p-k) (\gamma^\nu \gamma_5)^T] \; ,
\eeq
with
\beq
\label{6}     
   S(k) &\equiv& i\frac{\not\!k + M_c}{k^2-M_c^2} \; .
\eeq
Carrying out the traces and the $d^4 k$ momentum integral, one finds
\beq
\label{7}
 \Pi_{H}^{\mu \nu}(p) &=& 12 g^2[g^{\mu\nu}(M_c^2I_H(p)-
p_\alpha I_H^\alpha(p))+2p^\mu I_H^\nu(p)-2I_H^{\mu \nu}(p)] \;{\rm ,\;with} \\
  I_H(p) &=& \frac{2 M_c^2-p^2/2}{4 \pi^2} I_o(p) \nonumber \\
  I_H^\nu(p) &=& p^\nu[\frac{2 M_c^2-p^2/2}{8 \pi^2} I_o(p)+7/4] \nonumber \\
  I_H^{\mu \nu}(p)&=& \frac{g^{\mu\nu}}{4\cdot4 \pi^2 }([2M_c^4
-M_c^2 p^2/2 +(p^2/6)(5 M_c^2-p^2-4 M_c^4/p^2)] I_o(p)+constant)
\nonumber \\
 && +\frac{p^\mu p^\nu}{6\cdot4 \pi^2 }[(5 M_c^2-p^2-4 M_c^4/p^2)I_o(p)
-4 M_c^2/p^2] \nonumber \\
  I_o(p)&=& \int_0^1 \frac{d \alpha}{p^2(\alpha-\alpha^2) -M_c^2} 
\nonumber \; .
\eeq

As in our calculations using QCD sum rules\cite{lsk09,kpr08} we use the
scalar component, $\Pi_{H}^S$, defined as $ \Pi_{H}^{\mu \nu}(p)=
((p^\mu p^\nu/p^2)-g^{\mu \nu} )\Pi^V(p) + (p^\mu p^\nu/p^2)\Pi^S(p)$, with
$\Pi_{H}^V(p)$ the vector component. The vector component provides a separate
sum rule, but the sum rule with the scalar component has less error (as 
determined from the method of QCD sum rules in previous calculations-see, e.g.,
Refs\cite{kpr08,lsk09}). From 
Eq(6) we obtain
\beq
\label{8}
    \Pi_H^S(p)&=& \frac{12}{(4 \pi)^2}(8 M_c^4/3 -4 M_c^2 p^2/3 
+p^4/6)I_0(p) + constant + constant\cdot p^2 \; .
\eeq  
Using the Borel transforms from momentum to the Borel mass, $M_B$
\beq
\label{9}
     \mathcal{B}I_0(p) &=& 2 e^{-z} K_0(z) \nonumber \\
   \mathcal{B}p^2 I_0(p)&=& 4M_c^2e^{-z}[K_0(z)+K_1(z)] \nonumber \\
   \mathcal{B}p^4 I_0(p)&=& 2M_c^4e^{-z}[6 K_0(z)+8 K_1(z)+2 K_2(z)]
\eeq
with $z=2 M_c^2/M_B^2$ and $K_n$ are modified Bessel functions.
Therefore we find 
\beq
\label{10}
    \Pi_H^S(M_B)&=&  \frac{4 M_c^4}{(4 \pi)^2}g^2 e^{-z}[6 K_0(z) -8 K_1(z)
+2 K_2(z)]
\eeq. 
 
To find the two-pion decay of the $\Psi'(2S)$ we use the external field 
method\cite{cpw85,hhk96} with the quark propagating with an external pion, 
$S_\pi(p)$, as shown in Fig 5(a),
\beq
\label{11}
    S_\pi(p) &=& i g_\pi \tau \cdot \phi_\pi \frac{1}{\not\!p -M} \gamma_5
\frac{1}{\not\!p -M} \; ,
\eeq
where $g_\pi$ is the pion-quark coupling constant and $\phi_\pi$ is the pion
field.
\begin{figure}[ht]
\begin{center}
\epsfig{file=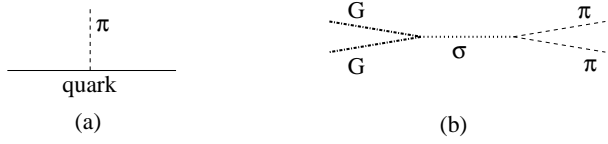,height=2cm,width=8cm}
\caption{ (a) Pion coupled to quark (b) Sigma coupled to gluon decays
to two pions}
\label{Fig.5}
\end{center}
\end{figure}

  From this we obtain the two-pion decay of the normal component of 
$\Psi'(2S)$, corresponding to Fig. 1 as
\beq
\label{12}
  \Pi_{H \pi \pi}^{\mu \nu} &=& -3 g_\pi^2 g^2 \int \frac{d^4 k}{(4 \pi)^4} 
\frac{Tr[\gamma^\mu \gamma^5 \gamma^\nu \gamma^5]}{(k^2-M_c^2)
((p-k)^2-M_c^2)} \nonumber \\
   &=& \frac{12}{(4 \pi)^2}g^{\mu \nu}g_\pi^2 g^2[(2 M_c^2-p^2/2)I_0(p)+
constant] \; ,
\eeq
from which one obtains, after a Borel transform, for the scalar 
component of the two-pion correlator
\beq
\label{13}
 \Pi_{H \pi \pi}^S(M_B) &=& \frac{24}{(4 \pi)^2}g_\pi^2 g^2 M_c^2
e^{-z}[K_0(z)-K_1(z)] \; .
\eeq

  The correlator $\Pi_{H-HH \pi \pi}^S(M_B)$ for $2 \pi$ decay from the
hybrid component of the $\Psi'(2S)$ to the $J/\Psi(1S)$ state, obtained 
from the diagram in Fig.2, turns out to be less than 1 \% of 
$\Pi_{H \pi \pi}^{S}(M_B)$ of Fig. 1 and is not included.

   Corresponding to the diagram shown in Fig.3, the correlator for the 
hybrid component of  $\Psi'(2S)$ to decay to
a sigma and the $J/\Psi(1S)$ state,  $\Pi_{H-HH \sigma}^{\mu\nu}(p)$, is 
obtained using two results from
previous work. First, it was shown in Ref\cite{lsk09} that the coupling
of the mixed to the normal charmonium, $\Pi_{H-HH}^S(M_B)$,  shown in 
Fig. 4(b), is given approximately by
\beq
\label{14}
 \Pi_{H-HH}^S(M_B) & \simeq & \pi^2 \Pi_H^S(M_B) \; ,
\eeq
where $\Pi_H^S(M_B)$ is the coupling of the normal component of $\Psi'(2S)$ 
to the $J/\Psi(1S)$ state.

   Next we use the external field coupling shown in Fig.5(b) 
to relate  $\Pi_{H-HH \sigma}^{\mu\nu}(p)$ to the mixed coupling,
$\Pi_{H-HH}^S(M_B)$ which gives
\beq
\label{15}
  \Pi_{H-HH \sigma}^S &=& \frac{g_\sigma}{M_\sigma}<G^2>\Pi_{H-HH}^S
\nonumber \\
   &=& \frac{g_\sigma}{M_\sigma}<G^2> \pi^2 \Pi_H^S \; ,
\eeq
from which we find
\beq
\label{16}
  \Pi_{H-HH \sigma}^S(M_B) &=&  \frac{24}{(4 \pi)^2}g^2
 \frac{g_\sigma}{M_\sigma}<G^2> M_c^2 \pi^2 e^{-z}[ K_0(z)
-4 K_1(z)/3+ K_2(z)/3] \; ,
\eeq
where $g_\sigma$ is the sigma-gluon coupling constant and $<G^2>$ is
the gluon condensate.

  From Eqs(13,16), with b=$1/\sqrt{2}$ we obtain R = ratio of sigma to 
free two-pion (excluding the sigma) decay widths as
\beq
\label{17}
   R&=& N \frac{\frac{g_\sigma}{M_\sigma}<G^2> M_c^2 \pi^2[ K_0(z)
-4 K_1(z)/3+ K_2(z)/3]}{g_\pi^2[K_0(z)-K_1(z)]} \; ,
\eeq
with N a normalization factor derived below.  The gluon-sigma coupling 
constant, $g_\sigma$, was found using a low-energy theorem\cite{nov80} and 
has been used successfully in studying scalar glueballs\cite{kj01}, the 
production of sigmas in high-energy proton-proton
collisions\cite{kms05}, and the decay of hybrid baryons\cite{kl99}.
The result is $g_\sigma/M_\sigma \simeq 1.0$. 
The pion quark coupling is known to be\cite{DZ86} 
$g_\pi=f_\pi m_\pi^2/(m_u + m_d)$.
Using standard values for the quark masses, $m_u = m_d$ = 3 MeV, 
$g_\pi^2 = 0.189$ $GeV^4$. The gluon condensate $<G^2>=0.476 GeV^4$. 
\vspace{1cm}

\begin{figure}[ht]
\begin{center}
\epsfig{file=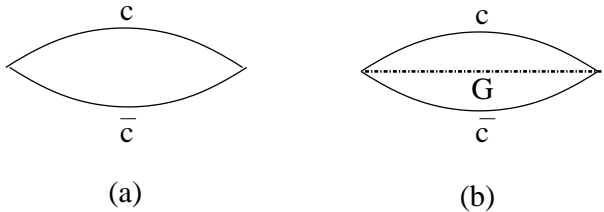,height=2cm,width=8cm}
\caption{ (a) Normal charmonium normalization (b) Hybrid charmonium
normalization}
\label{Fig.6}
\end{center}
\end{figure}  

  Finally, since the correlators do not use normalized states we must
find the relative normalization of the normal and hybrid components.
The normalization parameter is found from
\beq
\label{18}
   N^2 &=& \frac{\int d(M_B^2) \Pi_H^S(M_B)}{\int d(M_B^2)
 \Pi_{H-HH}^S(M_B)} \; .
\eeq

Using Eq(10) for $\Pi_H^S(M_B)$ and Eq(15) in Ref\cite{kpr08} for
$\Pi_{H-HH}^S(M_B)$ we find that N=0.0123 $M_c^2$.
Using the charmonium mass, $M_c$, and the Borel mass, $M_B$=mass of
the $\Psi'(2S)$, and ($K_0(z), K_1(z), K_2(z)$=(1.54, 3.75, 31.53), we 
find for $R$ = ratio of $\sigma$ to $2 \pi$ decay of the $\Psi'(2S)$ to 
the $J/\Psi$
\beq
\label{19}
              R&=&0.98 \; .
\eeq

This result is in very good agreement with recent measurements by the
BES Collaboration at the IHEP, Beijing, \cite{BES07}. The uncertainty
in the $R$ is not large, similar to the rapid convergence of diagrams
used in QCD sum rules for heavy quark states. Since we are using only
the lowest order diagram, from these previous calculations
we estimate an uncertainty of about 15 percent, which was shown to arise
mainly from the continuum contributions. This estimate of error is based
on the method of QCD sum rules 

\section{Conclusion}

Using the results of our previous study that found the $\Psi'(2S)$ to
be a mixed normal and hybrid charmonium state, we have found the branching
ratio of the sigma to two-pion decays of this state to the $J/\Psi$ to
be in good agreement with BES measurements. With the successful application
of this theory to explain the so-called $\rho-\pi$ decay problem of
the $\Psi'(2S)$\cite{lsk09}, this gives further evidence that the $\Psi'(2S)$
is a mixed heavy-quark hybrid. In future studies we shall explore other
decays, and the production of the $\Psi'(2S)$ in Relativistic Heavy Ion
Collisions, where the octet model has been shown to be dominant\cite{nlc03,
cln04}, consistent with the color 8 gluon of the hybrid component.

\section{Acknowledgements}

This work was supported in part be the NSF/INT grant number 0529828. The
authors thank both theorists and experimentalists at the IHEP, Beijing 
for helpful discussions.

\end{document}